\documentclass[
]{ceurart}

\usepackage{listings}
\usepackage{subcaption}
\usepackage{listings}
\usepackage{xcolor}
\usepackage{hyperref}
\begin{document}

\copyrightyear{2026}

\copyrightclause{Copyright for this paper by its authors. Use permitted under Creative Commons License Attribution 4.0 International (CC BY 4.0).}
\conference{Data-Driven Storytelling: Bridging Knowledge Graphs, GenAI, and Narrative (DDS2026), in ISWC 2026 Workshops Joint Proceedings, October 25--26, 2026, Bari, Italy}

\title{From Triples to Cards: A Web Component for Human-Readable Linked Open Data}


\author[1]{Puyu Wang}[%
orcid=0009-0003-8089-4146,
email=puyu.wang@eng.ox.ac.uk,
url=https://puyuwang.org/,
]
\cormark[1]
\address[1]{Oxford e-Research Centre, Department of Engineering Science, University of Oxford, Oxford, UK}

\author[2]{Hang Dong}[%
orcid=0000-0001-6828-6891,
email=h.dong2@exeter.ac.uk,
url=https://experts.exeter.ac.uk/41644-hang-dong,
]
\address[2]{Department of Computer Science, University of Exeter, Exeter, UK}

\author[3]{Rui Zhao}[%
orcid=0000-0003-2993-2023 ,
email=zhaorui@tju.edu.cn ,
url=https://me.ryey.icu,
]
\address[3]{School of Cybersecurity, Tianjin University, Tianjin, China}

\cortext[1]{Corresponding author.}

\begin{abstract}
JSON-LD is getting increasing attention for publishing Linked Data, especially within cultural heritage. We present LODox Data Viewer, a reusable web component library for data presentation and visualization in semantic portals. It renders JSON-LD as human-readable, themeable cards, embeddable in any semantic portal page with a single script tag. 

\textbf{Package URL:} https://www.npmjs.com/package/lodox-data-viewer
\end{abstract}


\maketitle

\section{Introduction}
JSON (JavaScript Object Notation) has become the dominant data format for web applications \cite{pezoa_foundations_2016,bray_javascript_2017}. JSON-LD, a Linked Data serialization format based on JSON \cite{kellogg_json-ld_2019}, is designed to ease the adoption and integration of Linked Data in web applications. It has been attracting increasing attention, especially in the cultural heritage domain. For example, the Linked Art data model \footnote{https://linked.art/api/1.0/json-ld/} provides a JSON-LD based data profile built on the CIDOC-CRM ontology \footnote{https://cidoc-crm.org}, aiming to provide shared data modelling approaches \cite{delmas-glass_fostering_2020}. It has been adopted and implemented in a range of online applications and workflows \cite{raemy_enabling_2023,sanderson_implementing_2024}. Furthermore, the Getty Museum Collection also provides an API based on Linked Art \footnote{https://data.getty.edu/museum/collection/docs/}. 

Meanwhile, many existing semantic portals and web applications for presenting cultural heritage Linked Data offer rich interfaces but rely on a full stack, so they cannot easily be integrated into other systems without deploying that complete stack \cite{vrandecic_reshaping_2018,doi:10.3233/SW-210428}. We developed \emph{LODox (Linked Open Data Oxford) Data Viewer}, a reusable web component library that is: 
\begin{itemize}
    \item \textbf{easily embedded}: it can be embedded anywhere with one script tag by different web applications. 
    \item \textbf{able to render JSON-LD without per-record templates}: the viewer renders JSON-LD data into human-readable cards. 
    \item \textbf{customizable}: per-record layout and theming can be configured.
\end{itemize}

On a further note, LODox Data Viewer is still under active development. It was initially the presentation layer of a full-stack system for publishing cultural heritage Linked Data, namely LODox, and was later decoupled into a stand-alone library so that it can be reused independently.

In this paper, we will first describe the design and implementation of LODox Data Viewer (Section \ref{sec:design}), and demonstrate its usage with JSON-LD and Linked Art (Section \ref{sec:Demo}). Finally, we will discuss the conclusion and future plans in Section \ref{sec:conclusion}.  


\section{Design and Implementation}
\label{sec:design}
LODox Data Viewer is available either as an npm (Node Package Manager) package or as a loadable
bundle from a Content Delivery Network (CDN). For example, it can be loaded using unpkg.com:
\begin{lstlisting}[language=HTML]
<script type="module" src="(*\href{https://unpkg.com/lodox-data-viewer}{https://unpkg.com/lodox-data-viewer}*)"></script>
\end{lstlisting}
\begin{figure}[htbp]
    \centering
    \includegraphics[width=1\textwidth]{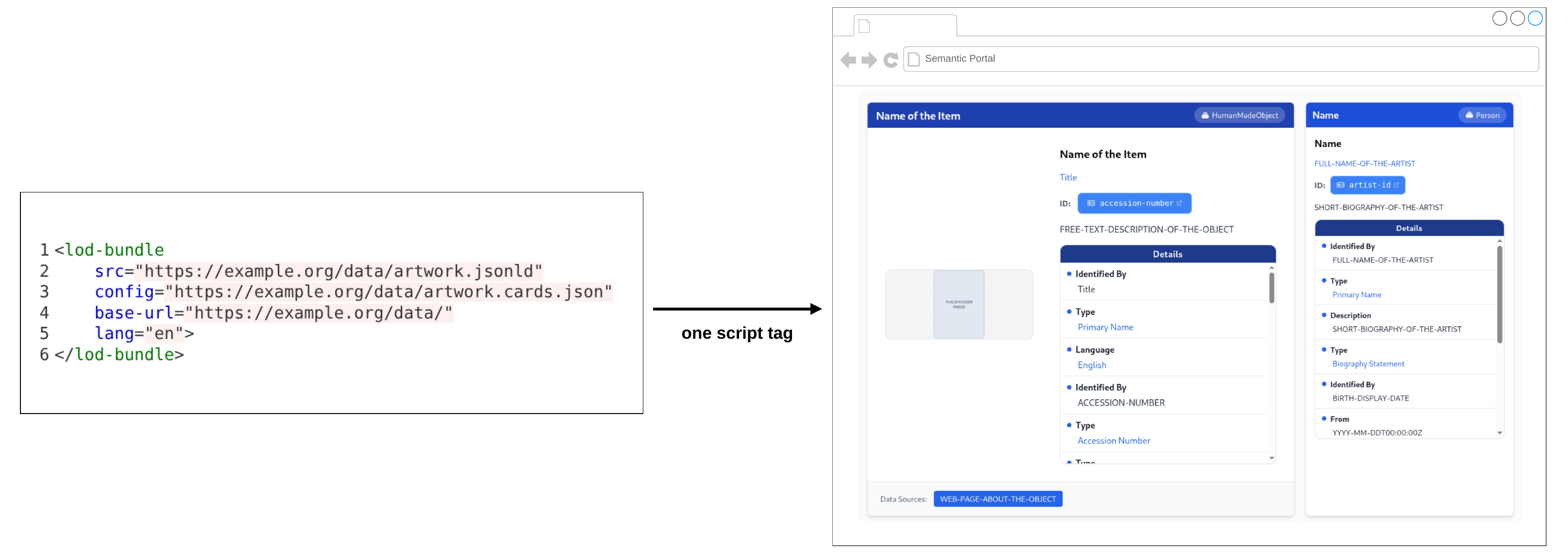}
    \caption{Rendering JSON-LD with a few lines of code}
    \label{fig:use}
\end{figure}

\begin{figure}[htbp]
    \centering
    \includegraphics[width=0.8\textwidth]{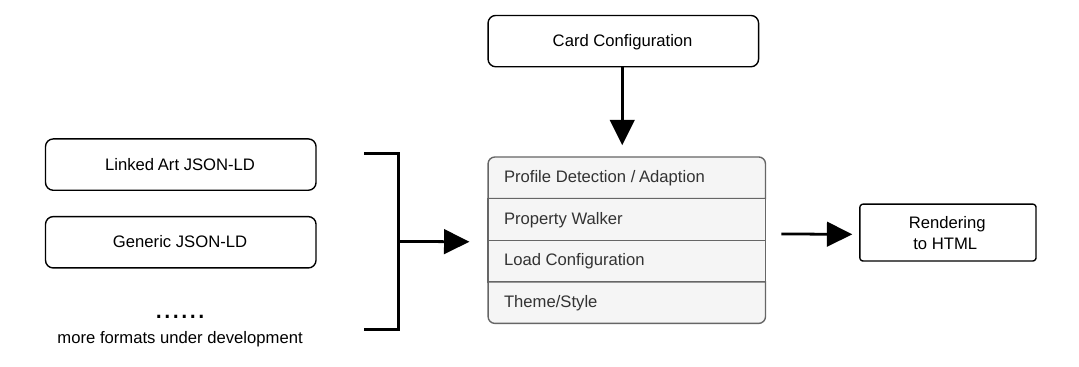}
    \caption{Render Workflow}
    \label{fig:workflow}
\end{figure}

From a few lines of code with a single script tag, LODox Data Viewer renders JSON-LD into multiple human-readable cards, as shown in Figure \ref{fig:use}. In order to do this, we use the workflow demonstrated in Figure \ref{fig:workflow}. First, the package automatically detects the profile of the input JSON-LD. At present, it detects the Linked Art profile and generic JSON-LD; support for additional profiles is under development. The corresponding adaptor decides what enters the card header as a record and extracts basic information for each record: identifier, link, title, description, image, etc. A \emph{walker} then traverses all the remaining properties losslessly: every non-empty value reachable from each record's root will appear on the corresponding record card, except for language tags and image captions. Nested objects are flattened. For example, measurement objects are represented as a single row containing both the measured value and its unit. An object with a \texttt{\_label}, a \texttt{content}, or an \texttt{id} emits a row of its own, followed by rows for its properties. An object without these emits no row of its own, but its properties may appear as rows. The layout, size, and colours of each card are configurable either by a configuration file or through CSS custom properties. The package also includes built-in presets, each setting a card's layout and size in a single word.

\begin{figure}[htbp]
    \centering
    \begin{subfigure}[t]{0.48\textwidth}
        \centering
        \includegraphics[width=\textwidth]{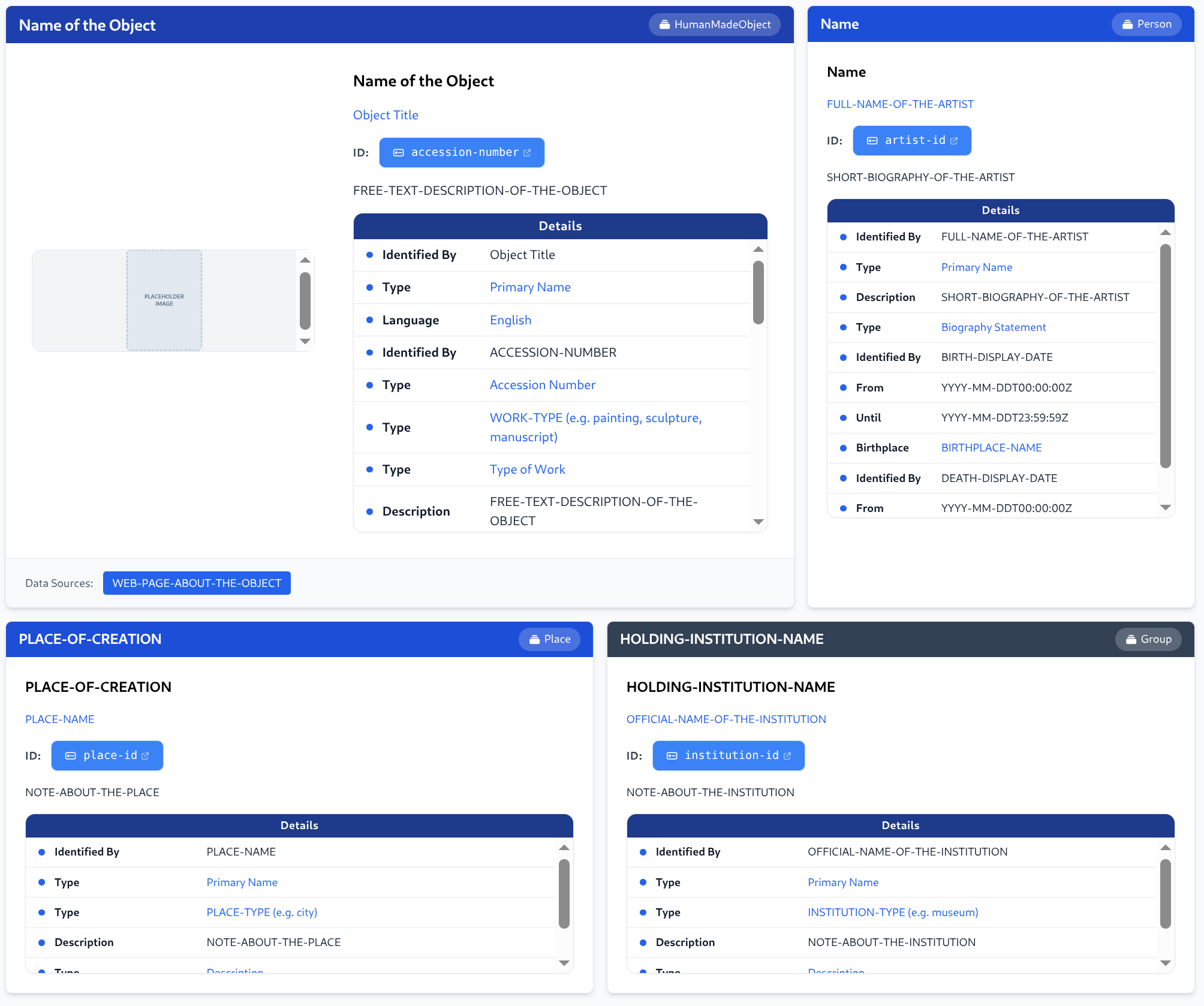}
        \caption{Multiple cards view}
        \label{fig:dataviewer_multi_card}
    \end{subfigure}
    \hfill
    \begin{subfigure}[t]{0.48\textwidth}
        \centering
        \includegraphics[width=\textwidth]{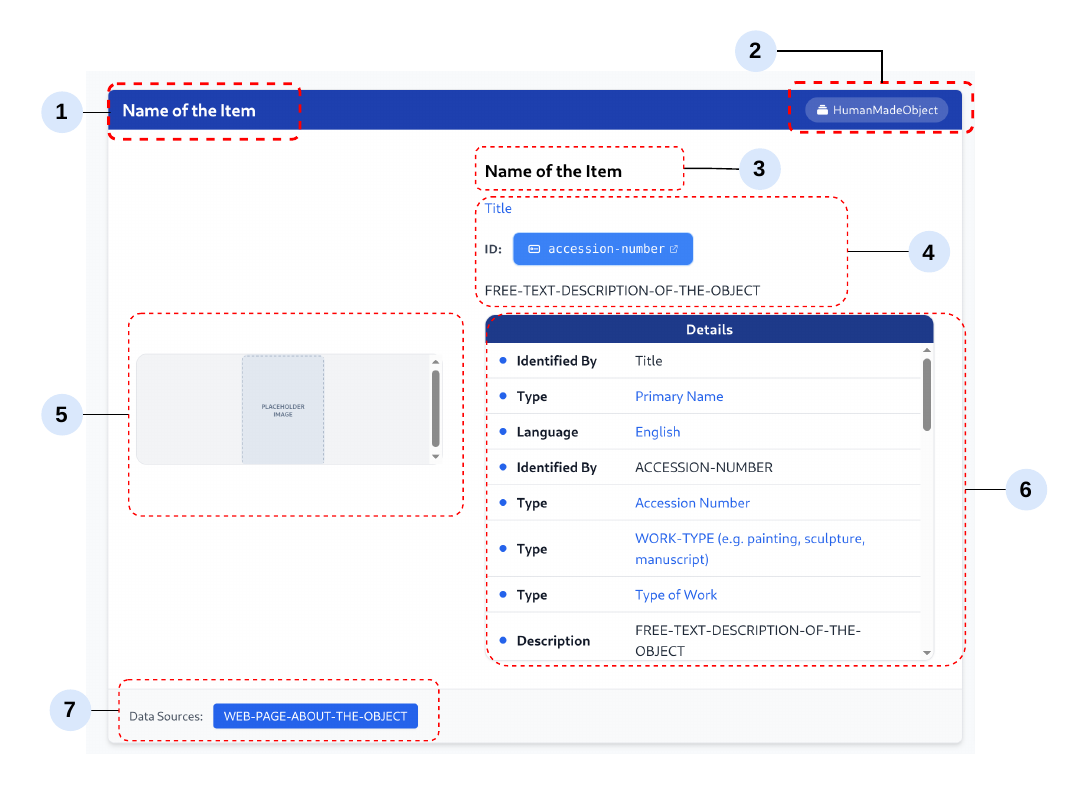}
        \caption{The card layout}
        \label{fig:card-detail}
    \end{subfigure}
    \caption{LODox Data Viewer interface, with the visualization of multiple cards and the layout of a card.}
    \label{fig:dataviewer_view}
\end{figure}

Finally, the package presents each record as a card including a header with the title and type, the images, a short description, and a table of detailed properties. A single JSON-LD file can contain a main item together with related ones, for example, a painting alongside its painter and current location. Each of these is rendered as its own card showing the properties relevant to it, as shown in Figure \ref{fig:dataviewer_multi_card}.

\begin{figure}[htbp]
    \centering
    \includegraphics[width=0.6
    \textwidth]{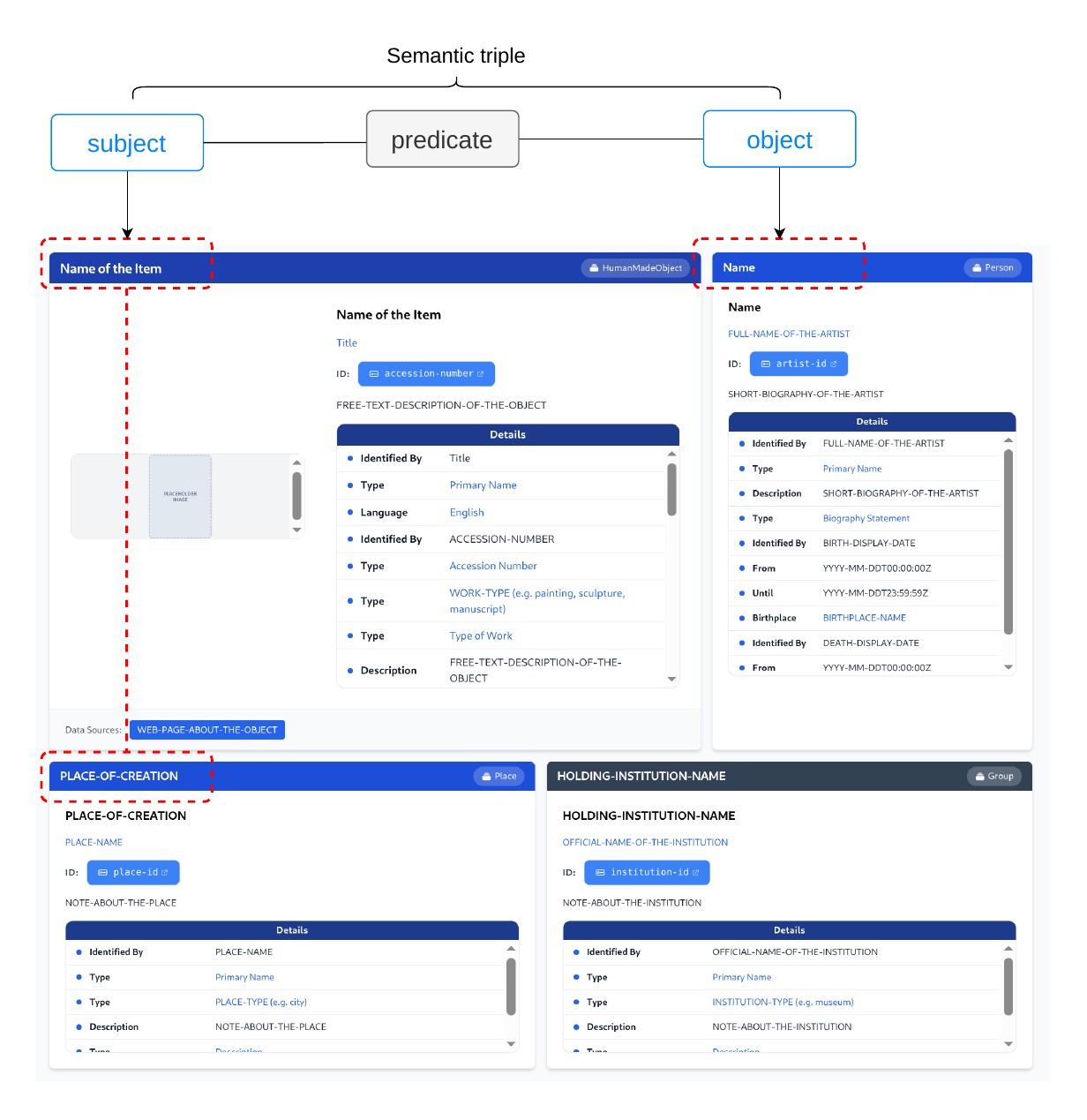}
    \caption{Multiple cards as triples, as visualized through the LODox Data Viewer interface.}
    \label{fig:muti-card}
\end{figure}

Figure \ref{fig:card-detail} shows the layout of a single card. (1) and (3) show the primary label, in the card header and at the top of the body. (2) is the object type. For Linked Art JSON-LD, it is the CIDOC-CRM class representing the resource type, for example, ``HumanMadeObject''\footnote{\url{http://www.cidoc-crm.org/cidoc-crm/E22_Human-Made_Object}}. (4) contains further essential information, including alternative names, the identifier (URI), and a short description. (5) contains the associated images. (6) contains a detailed table, which takes the record as the subject. Each row in the table shows one property and its value. Finally, (7), shown in the footer, displays the data source as a clickable link that directs users to the original source website. 
Furthermore, each card can be configured with different height and width, allowing the overall layout of the cards to be customized. The cards will be stacked side by side automatically given enough width, or vertically. Therefore, as the subject and object in a semantic triple can be placed as cards side by side, the layout can also be used to present relations between the cards, as shown in Figure \ref{fig:muti-card}. 

\section{Demonstration}
\label{sec:Demo}

\begin{figure}[h]
    \centering
    \includegraphics[width=1\textwidth]{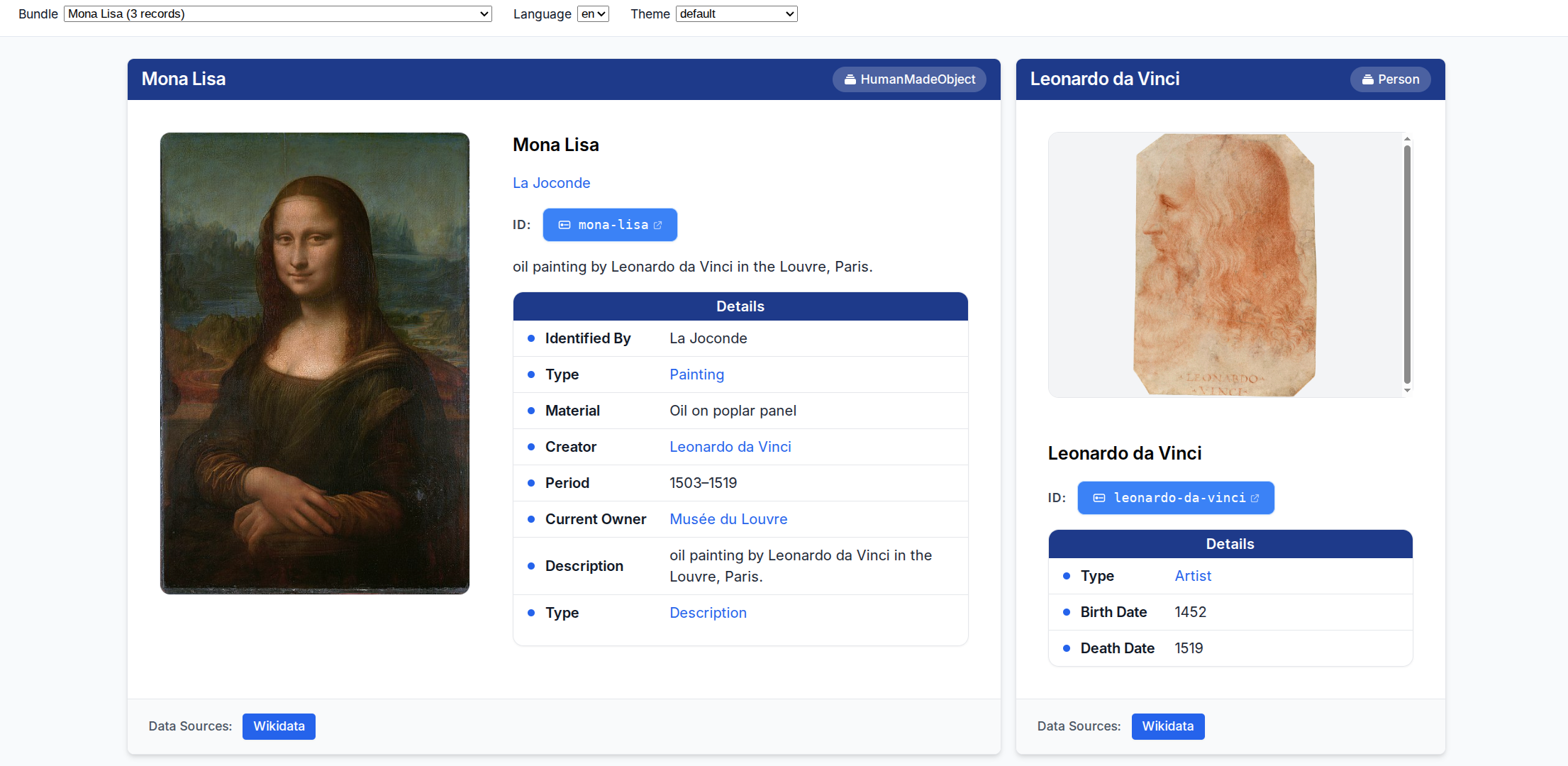}
    \caption{Screenshot from a live demo featuring Mona Lisa.}
    \label{fig:dataviewer}
\end{figure}

The npm package includes a command (\texttt{npm run demo}) that runs a live demonstration server locally. At present, Linked Art JSON-LD files are handled by an adaptor written for their vocabulary; others are rendered by the generic profile with no code specific to them. Figure~\ref{fig:dataviewer} shows the demo displaying the Mona Lisa together with Leonardo da Vinci\footnote{The JSON-LD data used in this example is available online: \url{https://github.com/cogim-online/lodox_card_demo/blob/main/data/mona-lisa.jsonld}}. 

\section{Conclusion and Future Work}
\label{sec:conclusion}

In this paper, we presented LODox Data Viewer, an npm package for rendering JSON-LD. We have demonstrated that JSON-LD data with well-established vocabularies can be effortlessly rendered while emphasizing the  fields specified by the profile adaptors. LODox Data Viewer also provides a card-based presentation for Linked Data. The package remains an early-stage software demonstration, and we have not yet conducted a user-based evaluation. The usability is evaluated based on development experience and unit testing. User-based evaluation with developers and practitioners is planned as future work. In the meantime, we identify three major directions for future development. 

First, we will develop more adaptors for other well-established ontologies and investigate how to render data described by multiple ontologies simultaneously. 

Second, we acknowledge that the rendering result depends on the configuration and the quality of the JSON-LD provided. At present, not every JSON-LD file renders well owing to limited profile compatibility and data completeness. We therefore plan to investigate algorithms and approaches for auto-configuration, enabling more flexible rendering. 

Third, we will investigate the use of AI agents to retrieve and control the cards and terms expressed in JSON-LD, in particular, to determine which cards are included in a given JSON-LD document. We expect LODox Data Viewer to be embedded in a pipeline in which an agent reviews data drawn from a graph database and exports the corresponding JSON-LD. Moreover, the current package only renders static JSON-LD files. We also wish to develop dynamic loading with an agentic AI, in which cards are generated successively, each one selected for its relevance or relations to those already selected.
\begin{acknowledgments}
The authors are grateful to Professor Donna Kurtz for her guidance and support for this work.
\end{acknowledgments}

\section*{Declaration on Generative AI}

 During the preparation of this work, the author(s) used Claude and ChatGPT in order to: grammar and spelling check, content enhancement, improve writing style, and paraphrase and reword. After using these tools/services, the authors reviewed and edited the content as needed and take full responsibility for the publication’s content. 

\bibliography{references}


\end{document}